          [2001/04/25 1.1 (PWD)]
\documentclass[a4paper,twocolumn]{esapub} % European paper

\usepackage{times}
\usepackage{natbib}
\usepackage{graphicx}

\def\aj{AJ}                   
             
\def\apj{ApJ}

\def\aap{A\&A}     
\def\mnras{MNRAS}             
\def\nat{Nature}              
              
\def\pasp{PASP}              
\def\iaucirc{IAU~Circ.}

\def\lsim{\!\!\!\phantom{\le}\smash{\buildrel{}\over
  {\lower2.5dd\hbox{$\buildrel{\lower2dd\hbox{$\displaystyle<$}}\over
                               \sim$}}}\,\,}
\def\gsim{\!\!\!\phantom{\ge}\smash{\buildrel{}\over
  {\lower2.5dd\hbox{$\buildrel{\lower2dd\hbox{$\displaystyle>$}}\over
                               \sim$}}}\,\,}

\title{Supernova Cosmology and the ESSENCE project}

\author[1,2]{Jesper Sollerman}

\affil[1]{DARK Cosmology Center, Niels Bohr Institute, University of Copenhagen, Denmark}
\affil[2]{Department of Astronomy, Stockholm University, Sweden, jesper@astro.su.se}

\author[3]{Claudio Aguilera} 
\affil[3]{Cerro Tololo Inter-American Observatory}

\author[4]{Andy Becker} 
\affil[4]{University of Washington} 

\author[5]{Stephane Blondin} 
\affil[5]{European Southern Observatory} 

\author[6]{Pete Challis} 
\affil[6]{Harvard-Smithsonian Center for Astrophysics}

\author[7]{Alejandro Clocchiatti} 
\affil[7]{Universidad Catolica de Chile}
 
\author[8]{Alex Filippenko} 
\affil[8]{University of California, Berkeley }

\author[8]{Ryan Foley}

\author[9]{Peter M. Garnavich} 
\affil[9]{Notre Dame University } 

\author[6]{Malcolm Hicken}

\author[8]{Saurabh Jha} 

\author[6]{Robert Kirshner} 

\author[9]{Kevin Krisciunas} 

\author[5]{Bruno Leibundgut} 

\author[8]{Weidong Li}

\author[10]{Thomas Matheson} 
\affil[10]{National Optical Astronomy Observatory} 

\author[4]{Gajus Miknaitis}

\author[3]{Armin Rest}

\author[11]{Adam G. Riess} 
\affil[11]{Space Telescope Science Institute } 

\author[12]{Maria Elena Salvo} 
\affil[12]{RSAA, The Australian National University }

\author[12]{Brian P. Schmidt}

\author[3]{Chris Smith}

\author[5]{Jason Spyromilio}

\author[4]{Chris Stubbs}

\author[3]{Nicholas B. Suntzeff}

\author[13]{John L. Tonry} 
\affil[13]{University of Hawaii } 

\author[6]{Michael Wood-Vasey}

\author[13]{Brian Barris}

\author[1,12]{Tamara Davis}

\author[2]{Edvard M\"ortsell}

\begin{document}


\maketitle

\begin{abstract}

The proper usage of Type Ia supernovae (SNe Ia) as distance indicators has revolutionized cosmology, and added a new dominant component to the energy density of the Universe, {\bf dark energy}.
Following the discovery and confirmation era, the currently ongoing SNe Ia surveys aim to determine the properties of the dark energy. ESSENCE is a five year ground-based supernova survey aimed at finding and characterizing 200 SNe Ia in the redshift domain $z=$[0.2$-$0.8]. The goal of the project is to put constraints on the equation of state parameter, $w$, of the dark energy with an accuracy of $\lsim10\%$. This paper presents these ongoing efforts in the context of the current developments in observational cosmology.
\end{abstract}

\section{Introduction}

Supernova measurements have profoundly changed cosmology. The first results to argue for an accelerated rate of cosmic expansion, and thus a repulsive dark energy component, have already matured for 7 years \citep{riess98,perlmutter99}.
Today these results are accommodated in what has become the concordance cosmology, flanked by constraints on the matter density, $\Omega_{\rm M}$, from large scale structure measurements, and on the flatness of space from CMB measurements.
This concordance cosmology is dominated by the dark energy, $\Omega_{\rm X}\simeq2/3$, and all present evidence is consistent with an interpretation of the dark energy as Einstein's cosmological constant, $\Lambda$ \citep{einstein17}.

From the supernova cosmology perspective, the years following the 1998 discovery focused to a large extent on confirming the early results with larger and independent supernova samples, and on further investigation of potential systematic uncertainties \citep[see e.g.,][for reviews]{leibundguts01,leibundgut01,filippenko04}. Within the high-z supernova search team (HZT), this effort culminated in 2003 with the analysis of over 200 Type Ia supernovae \citep{tonry03}. That work investigated a large number of potential pitfalls for using Type Ia supernovae in cosmology, but found none of them to be severe enough to threaten the conclusions of the 1998 paper. 

With 230 SNe Ia, whereof 79 at redshifts greater than 0.3, the Tonry et al. (2003) compilation already provided interesting constraints on the dark energy.
This dataset was further extended and investigated by the HZT in 
\citet{barris04} and was later also
adopted by \citet{riess04}, who added a few significant SNe Ia at higher redshifts. However, combining supernova data from a large variety of sources also raised many concerns, and it became increasingly evident that an improved attack on the $w$-parameter (Section 2) would require a systematic and coherent survey. Most of the members in the High-z supernova search team therefore climbed the next step, into the ESSENCE project (Section 3).

\section{The equation of state parameter}

Any component of the energy density in the Universe can be parameterized using a sort of equation-of-state parameter $w$, relating the pressure (P) to the density ($\rho$) via P~=~$w~\rho$~c$^{2}$. This parameter characterizes how the energy density evolves with the scale factor, $a$; 
$\rho~\propto~a^{-3(1+w)}$. In that sense, 
normal pressure-less matter ($w=0$) dilutes with the free expansion as $a^{-3}$, while a cosmological constant component with $w=-1$ always keeps the same energy density.

The very fact that the cosmic expansion is accelerating means that 
the average energy density has an equation of state parameter of $< -1/3$.
The first supernova constraints on 
the dark energy equation of state
by \citet{garnavich98} indicated $w < -0.6$ (95\% confidence, for a flat universe with $\Omega_{\rm M} > 0.1$), and the extended analysis by Tonry et al. (2003) dictates that
$-1.48 < w < -0.72$ (95\% confidence for a flat Universe and a prior on $\Omega_{\rm M}$ from the 2dFGRS).

It seems that all the current supernova measurements, as well as independent ways to estimate $w$, are consistent with a cosmological constant, $w=-1$ \citep[e.g.,][]{mortsell04}. But this is not an unproblematic conclusion. Although the modern version of $\Lambda$ can be interpreted as some kind of vacuum energy \citep[e.g.,][]{carroll01}, the magnitude of the dark energy density implied by the supernova measurements is ridiculously many orders of magnitudes larger than suggested by fundamental physics. It is also difficult to understand why we happen to live in an era when 
 $\Omega_{\Lambda}$ and  $\Omega_{\rm M}$ are almost equal.

Given these objections against the cosmological constant, a variety of suggestions for new physics have emerged. Many models use evolving scalar fields, so called quintessence models \citep[e.g.,][]{caldwell98}, which allow a time-varying equation of state to track the matter density. In such models, the time averaged absolute value of $w$ is likely to differ from unity.
Many other models including all kind of exotica are on the market, like k-essence, domain walls, frustrated topological defects and extra dimensions. All of these, and even some versions of modified gravity models, can be parameterized using $w$.

An attempt to actually quantify the dark energy could therefore aim at determining $w$ to a higher degree of precision. The project ESSENCE is designed to determine $w$ to an accuracy of $\pm10\%$.
With that, we hope to answer one simple but important question; is the value of $w$ consistent with $-1$?

\section{The ESSENCE project}

The ESSENCE (Equation of State: SupErNovae trace Cosmic Expansion) project is a 5 year ground-based survey designed to detect and follow 200 SNe Ia in the redshift range $z=[0.2-0.8]$. 

\subsection{Strategy}

Finding and following large batches of distant supernovae has almost become routine operation. The first heroic attempt by \citet{danes89} is replaced with modern wide field cameras using large CCDs and automatic pipelines for real-time object detection. As mentioned above, uniform data is required for precision measurements, and ESSENCE is therefore acquiring all photometric data with the same telescope and instrument.

Given the available telescope time, we have performed Monte Carlo simulations to optimize the constraints on $w$ from our supernova survey
\citep{miknaitis03,miknaitis06}. The optimal strategy favors maximizing the area imaged, i.e., it is more efficient to monitor a large field with many SNe Ia, compared to a deeper study of a narrower field to reach a few more $z>0.7$ SNe.

In order to reach our goal (Sect.~\ref{goal}) we will need $\sim200$ well measured SNe Ia distributed evenly over the targeted redshift range. That this is a very efficient way to constrain $w$ was shown by \citet{huterer01}.

In principle, the best independent supernova probe of cosmology needs to use a wide redshift distribution, in order to break the degeneracy between 
$\Omega_{\rm M}$ and $\Omega_{\rm X}$ \citep[e.g.,][]{goobar95}. Future space-based supernova surveys will do so. But given the precise $\Omega_{\rm M}$ measurements already available within the concordance cosmology, a ground based supernova survey may exchange some of the more expensive $z>1$ SNe with a prior on the matter density. This is how the ESSENCE project works.

\begin{figure}
\centering
\includegraphics[width=1.1\linewidth]{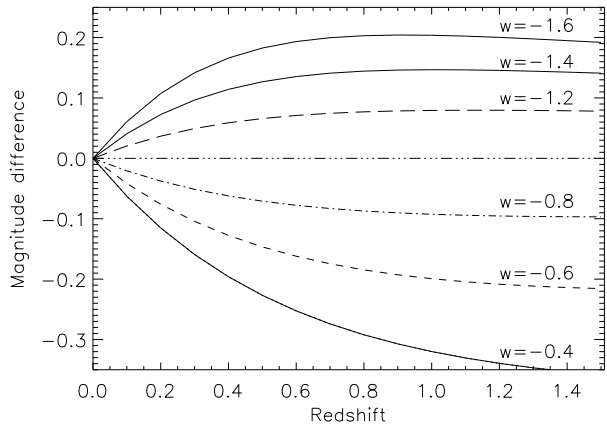}
\caption{Predicted difference in luminosity for world models with different values for $w$. All models have been calculated with the same cosmological parameters ($\Omega_{\rm M}, \Omega_{\rm X}$, H$_{0}$)
and are here compared to the value for the cosmological constant $w=-1$.
Even at the moderate redshifts targeted in the ESSENCE project, a measureable difference in the luminosity distances is predicted.
\label{f:wplot}}
\end{figure}

The interesting aspect for a supernova project is that a sizeable effect of the equation-of-state parameter can already be seen at the moderate redshifts where a ground-based survey is feasible. In Fig.~\ref{f:wplot} we show the differences in world models calculated for different values of $w$. All these models have used the same cosmology ($\Omega_{\rm M}=0.3, \Omega_{\rm X}=0.7$, H$_{0}=72$~km~s$^{-1}$~Mpc$^{-1}$), and the figure shows the expected magnitude differences as compared to a  $w=-1$ model. We see that there is already appreciable signal at redshifts around $z=0.5$. 
This is the motivation behind the ESSENCE project.

%\begin{small}
%\begin{verbatim}
%\end{verbatim}
%\end{small}

We will populate every $\delta z=0.1$ bin on the Hubble diagram with $>30$ SNe, and thus decrease the intrinsic scatter ($<0.14$~mag) to $\sim2.5\%$ uncertainty in distance modulus.
%print,0.14/sqrt(200/7.)    0.0261916
This, we believe (Sect.~\ref{goal}) is similar to our systematic uncertainties, and would, together with a $0.1 (1\sigma)$ fractional uncertainty on $\Omega_{\rm M}$ provide the required accuracy of the $w$-determination.
% 0.03 mag is dominated by K-corrections
From the predictions in Fig.~\ref{f:wplot} we note that 
at  $z=0.6$ the difference in luminosity models between $w=-1$ and $w=-1.1$ is 0.038 magnitudes.

\subsection{Implementation}

The work horse for the ESSENCE survey is the Blanco 4m telescope at CTIO, equipped with the Mosaic II Imager. The field-of-view for this imager is $36\times36
$ arc-minutes.
For the 5 year duration of this endeavor, we will
observe every second night during dark and dark/grey time for three consecutive months each Northern fall. We follow 32 fields that are distributed close to the celestial equator, so that they can be reached by (large) telescopes from both hemispheres. 
These fields were selected to have low galactic extinction, be free from very bright stars, and be located away from the galactic (and ecliptic) plane.
Furthermore, the distribution in RA of the fields must allow observations at reasonable airmass over the entire semester. The total sky coverage of the search is thus 11.5 square-arcminutes.
%IDL> print,(36./60)^2*32.      11.5200
The main part of the programme is to image each field in the $R$ and $I$ filter bands every 4 nights. This cadence allows us to detect the supernovae well before maximum light, and to simultaneously monitor the supernovae with enough sampling for accurate light-curve fits \citep[see e.g.,][]{krisciunas06}.

The pipeline automatically reduces the data and performs image subtraction. The software also rejects many artefacts, such as cosmic rays, as well as asteroids and UFOs. All remaining identified variable objects are potential SNe Ia, and are prioritized based on a rather complex set of selection criteria \citep{matheson05}. Spectroscopy is secured on the 8m class telescopes, such as the ESO VLT, Gemini, Magellan and the KECK telescopes. These spectra are used to (i) determine the redshift required to put the object onto the Hubble diagram, (ii) ensure that the object is a SN Ia, (iii) allow detailed comparisons between low-z and high-z supernova to look for evolution \citep[e.g.,][]{blondin06} and sometimes (iv) to derive an age estimate for the supernovae by comparison to local SN spectra.

It should be emphasized that the usage of 8m telescopes has substantially improved the quality of the high-z supernova spectra \citep{leibundguts01,matheson05} as compared to the SNe Ia used for the original 1998-claims. 

Apart from the core-programme, the ESSENCE team and its members also embark on many complementary programmes to assess specific scientific issues related to the ESSENCE scientific goals. We have used the HST to study in detail several of the highest redshift SNe Ia in the ESSENCE sample \citep{krisciunas06} and have been allocated SPITZER observing time to study a small sub-sample of ESSENCE SNe also in the rest-frame K-band, where dust and evolution are likely to be less important.
There are also ongoing investigations to study e.g., ESSENCE host galaxies, time-dilation from ESSENCE spectra, and reddening constraints from additional Z-band imaging.

\subsection{Current status - three out of five seasons}\label{s:current}

%\begin{small}
%\begin{verbatim}
\begin{figure}
\centering
\includegraphics[width=1.1\linewidth]{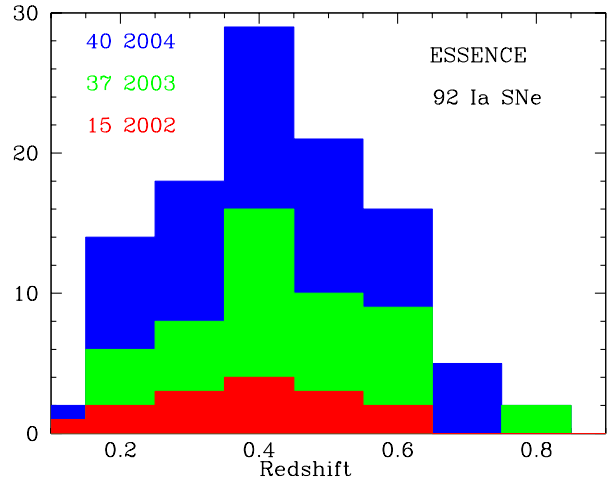}
\caption{The redshift distribution for the SNe Ia 
discovered by the ESSENCE project in the first 3 years.
\label{f:redshifts}}
\end{figure}
%\end{verbatim}
%\end{small}

We have now (summer 2005) finished three of the projected five years of the survey. We have detected about 100 SNe Ia (Fig.~\ref{f:redshifts}). All variable objects that we discover are immediately announced on the 
web\footnote{http://www.ctio.noao.edu/$\sim$wsne/index.html},
and the supernovae discovered by ESSENCE are announced in IAU circulars \citep{c02I,m02I,s02I,c03I,co03aI,co03bI,f04I,h04I,b05I}. 
We emphasize that all the images taken by the ESSENCE project are made public without further notice. Any researchers who could utilize such a uniform dataset for variable objects are welcome to do so.

The first ESSENCE paper described the spectroscopic part of the campaign \citep{matheson05}, and we have also discussed the properties of these spectra as compared to low-z supernovae \citep{blondin06} based on a newly developed optimal extraction method  \citep{blondin05}. The photometry for the nine supernovae monitored as part of our HST project has also been published \citep{krisciunas06}.

Overall, the project progresses as planned. The first season had a too low discovery rate of SNe Ia. This was largely due to bad weather, but we have also been able to improve the supernova finding software and to sharpen our selection criteria for spectroscopic follow-up, which means that the rates are now on track for the goal of 200 SNe Ia (Fig.~\ref{f:redshifts}).

Much of the work within the ESSENCE project has to date  been put on securing the observations and constructing the real-time data analysis system. At the moment, most of the efforts are put into the investigation of the systematic errors.
Different sub-groups of the team are working on e.g.;

(i) Photometric zero-point corrections %(0.9m calibrations, SDSS) 

(ii)  Redshift errors (from SN templates)

(iii) K-corrections (local spectral catalogue)

(iv)  Light curve shape corrections (different methods)

(v) Extinction law variations and Galactic extinction uncertainties

(vi) Selection effects, including Malmqvist Bias \citep{krisciunas06}.

It is also important to understand exactly how these different sources of uncertainties interact. They are clearly strongly correlated, and a robust error analysis technique that contains all these steps is required. \citet{krisciunas06} showed that light curve fits using three different methods were consistent with each other (their tables 4,5,6). But this comparison also showed some rather large differences for individual supernovae, which may require further investigation.

\subsection{Projected goal}\label{goal}

The aim of the project is to determine $w$ to $\pm0.1 (1\sigma)$. This is to be done by populating the Hubble diagram with a set of well observed SNe Ia in the redshift domain where we can probe the onset of the cosmic acceleration. This test is designed to examine whether or not this onset is consistent with the equation of state parameter of the cosmological constant. While it is of course of interest to also probe the time-evolution of a cosmological constant, this is very likely beyond the scope for the ESSENCE survey. Our constraints will thus be for the time-averaged value of $w$.

To be able to constrain the equation of state parameter $w$ to better than $\pm10\%$, we estimate that we need 200 SNe Ia to populate the Hubble diagram. How good the constraints will actually be will also depend on the adopted priors from other investigations.

For example, \citet{mortsell05} simulated the usage of 200 SNe with an intrinsic distance error of 0.14 mag, and distributed them over the anticipated ESSENCE redshift interval. We also added the 157 gold supernovae from \citet{riess04} as well as 300 local supernovae, as will be delivered by the SN factory \citep{aldering} or by the many other supernova searches conducted today, many of them including ESSENCE members \citep[e.g.,][]{li03,krisciunas04,jha05}.

\begin{figure}
\centering
\includegraphics[width=1.1\linewidth]{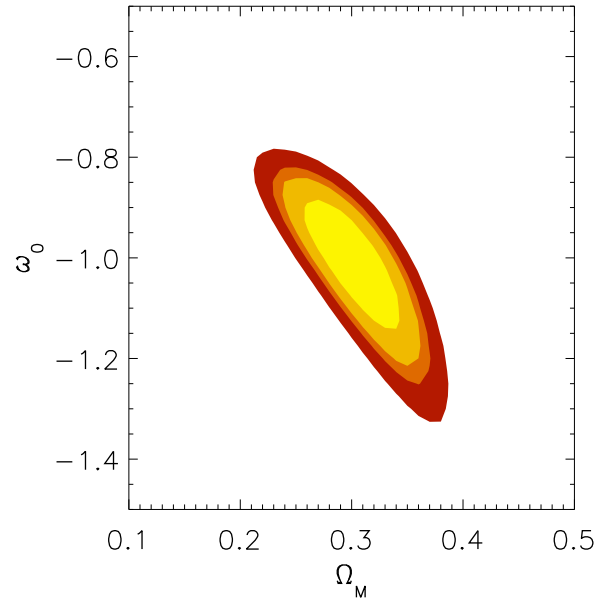}
\caption{Predicted constraints from future supernova studies from \citet{mortsell05}. These constraints are for 300 local SNe, the 157 gold supernovae and for 200 ESSENCE supernovae. In this plot we have also adopted a prior of 
$\Omega_{\rm M}=0.3\pm0.03$. 
A flat universe with a constant $w$-parameter is assumed. 
\label{f:edvardplot}}
\end{figure}

The anticipated constrains from this simulation 
are displayed in Fig.~\ref{f:edvardplot}.
If we furthermore adopt a conservative prior on $\Omega_{\rm M}$ of 10$\%$ we obtain a formal 1$\sigma$ error on a $w$ determination of $6-7\%$. This is as good as it gets.

The constraints will also depend on the systematic errors. These are more difficult to estimate, in particular prior to the actual experiment. It is likely that the battle with the systematics will be the most important one in this supernova survey.  
Many of the identified systematic uncertainties were listed in Sect.~\ref{s:current} and our pre-experiment estimates of the systematic floor is at the $2-3\%$ level. Thus, the survey is designed to reach the break-even point between systematic and statistical errors.

It can be of interest to compare the above-mentioned numbers with the other ambitious SNe Ia survey presently ongoing. The CFHT Supernova Legacy Survey 
(SNLS, Pain et al. 2003 and these proceedings)
aims to detect over 700 SNe Ia over the project lifetime. This is a substantial effort - not the least on the spectroscopic resources - where copious amounts of 8m class telescope time are required to identify all the candidates.

The first preliminary reports from the SNLS, based on the first year data only, appears to be very encouraging 
%(Astier et al. 2006, 
(Pain et al. 2006 these proceedings). Their error bar on $w$ is already as good as $10-11\%$ (RMS), including a systematic uncertainty of about half that amount. This is based on about 70 high-z supernovae.
If we assume that increasing the sample to 200 SNe will decrease the statistical noise by the Poisson contribution, the statistical error will become exactly equal to the quoted systematic error and the RMS error will be decreased to 
$\lsim0.08$.
But increasing the sample up to 700 supernovae would only extrapolate to an improvement to 
$\simeq0.06$
in the combined error. 

That the floor of the systematic error is likely to limit the experiment rather than the number of supernovae was the main consideration in limiting the ESSENCE survey to 200 SNe. To what extent the systematics can actually be better controlled, with or without a larger sample, will therefore determine the success of these surveys.

\section{Caveats}

Any supernova cosmology review will have to carefully mention the potential pitfalls in this game, including extinction, gravitational lensing, supernova evolution coupled to metallicity or other population effects as well as selection biases. Here we briefly mention the most obvious of these.

\subsection{Extinction}

Dimming by dust is always present in astronomy, although there is little evidence that this is severely affecting the SNe Ia cosmology \citep{riess98}. 
\citet{sullivan03} showed that the dark energy dominated cosmology persists even if only supernovae in elliptical galaxies are used, excluding strong bias due to local dust. Even models of grey intergalactic dust have been proposed, but seem to have fallen out of fashion.

\subsection{Luminosity Evolution}

Luminosity evolution was historically the major caveat in pinning down the deceleration parameter using e.g., (first-ranked) galaxies as standard candles. It is at least clear the SNe Ia do an enormously better job as standard(izable) candles. Empirically, many investigations have searched for luminosity differences depending on host galaxy type and redshift, but after light curve shape corrections no such differences have (yet) been found \citep[see e.g.,][and references therein]{filippenko04,gallanger05}.

In this respect we would of course feel much more confident if the theoretical backing of the SNe Ia phenomenon could further support the lack of evolution with redshift and/or metallicity.

The general text-book scenario for a SN Ia explosion is quite accepted; a degenerate carbon-oxygen white dwarf accreting matter by a companion star until it reaches the Chandrasekhar limit and explodes (at least initially) via deflagration. This thermonuclear blast completely disrupts the white dwarf, and converts a significant fraction of the mass to radioactive $^{56}$Ni, which powers the optical light curve.  But it is possible to take a more cautious viewpoint, since we have still not observed a single SN Ia progenitor white dwarf before it exploded, and in particular the nature of the companion star is hitherto unknown. It is quite possible that a multitude of progenitor system channels exists, and the redshift distributions of such populations are not known. Studies to detect and constrain the progenitor systems are ongoing, by e.g., investigating the present white dwarf binary population \citep{napi02} and by searching for circumstellar material at the explosion sites \citep[e.g.,][]{mattila05}.

Also the explosion models have developed significantly in recent years. R\"opke et al. (2005, these proceedings) present exploding 3D-models based on reasonable deflagration physics. But it is important to go beyond the simplest observables, the fact that the simulations should indeed explode with a decent amount of bang, and to compare the explosion models to real SNe Ia observations. An important step in this direction was made by \citet{kozma05} who modeled also the nucleosynthesis and the late spectral synthesis for comparison to optically thin nebular SNe Ia spectra. 
%\citep[e.g.,][]{sollerman04}. 
This initial attempt revealed the explosion models to produce far too much central oxygen, thus showing that efficient constraints can be directly put on the explosion models from properly selected observables. Hopefully, explosion models will soon converge to the state where it becomes possible to test to what extent a change in pre-explosion conditions - as may be 
suspected by altering metallicity or progenitor populations - will indeed affect the SNe Ia as standard candles.

An empirical way to investigate any potential redshift evolution is to compare the observables of the low redshift sample with those of the high redshift sample. The most detailed information is certainly available in the spectra, and 
\citet{blondin06} have used the ESSENCE spectra for such a detailed comparison. The main conclusion of that investigation is that no significant differences in line profile morphologies between the local and distant samples could be detected.

\subsection{Gravitational Lensing}

Gravitational lensing is also a potential concern. Present studies indicates that the effects are small at the redshift ranges populated by the ESSENCE supernovae, but that corrections could be made for higher redshift domains, as may be reached by JDEM/SNAP \citep{gunnarsson06,jonsson06}. 

\citet{jonsson06} recently modeled the lensing effect of 14 high-z SNe in the gold sample. %\citep{\riess04} . 
The original 157 SNe in that sample \citep{riess04} gives 
$\Omega_{\rm M}=0.32^{+0.04}_{-0.05}~(1\sigma)$
in a flat universe. If corrected for the foreground lensing as estimated by \citet{jonsson06}, the constraints instead becomes 
$\Omega_{\rm M}=0.30^{+0.06}_{-0.04}~(1\sigma)$
in a flat universe. This difference is indeed very small. 
There is no significant correlation between the magnification corrections and the residuals in the supernova Hubble diagram for the concordance cosmology.

\subsection{Selection Bias}

The selection of the SNe Ia followed by ESSENCE is far from homogeneous \citep{matheson05}. To decide which objects are most likely young SNe Ia candidates in the targeted redshift domain, and also suitable for spectroscopic identification and redshift determination, involves a complicated set of selection criteria. 
The final list also depends on the availability of spectroscopic telescope time. 
This may mean that the selection of the distant sample is different from the nearby, for example by favoring SNe placed far from the host galaxy nucleus. 

An argument against severe effects from such a bias in the distant sample is that also the nearby SNe Ia population - which is indeed shown to be excellent standard candles - is drawn from a large variety of host galaxies and environments.
There is therefore reason to believe that, as long as the physics is the same, the methods to correct for the reddening and light curve shape also holds for the high-z sample \citep[e.g.,][]{filippenko04}. In fact, in terms of cosmological evolution, the galaxies at $z\sim0.5$, where the supernova dark energy signal is strongest, have not evolved much. 

In \citet{krisciunas06} we show that in the high-z tail of the ESSENCE redshift distribution, we are susceptible for Malmqvist bias. This is the sample selected for follow-up with the Hubble Space Telescope. However, most of our survey is deep enough to be immune to this effect. Since we do need the light curve corrections all our supernovae have to be easily detected at maximum.

\subsection{Caveats - current status}

The above subsections have focused on the systematic uncertainties  
of the supernovae as standard candles throughout the universe. 
After the observations of $z\gsim1$ SNe, first hinted by \citet{tonry03}, but clearly detected by \citet{riess04}, much of the old worries about these uncertainties have disappeared. That the very distant supernovae are {\it brighter} than expected in a coasting universe, while the $z\sim0.5$ SNe are {\it fainter} than expected, is a tell-tale signal that rules out most reasonable dust or evolution scenarios. For sure, these kind of models can still logically be constructed - but must generally be regarded as contrived.

While the conclusions from all investigations hitherto conducted give reasonable confidence that none of the known caveats (alone) are serious enough to alter any of the published conclusions, the ongoing large surveys, and in particular any future space based missions, still have to seriously investigate these effects.
Clearly, the ESSENCE sample of well measured SNe Ia will make many of the requested tests for systematic effects possible to a much higher degree than hitherto possible. 
%As stated many times already, it will be important to beat down also these systematic uncertainties in order to best utilize this large dataset.

\section{Discussion}

Astronomers coming from the supernova field always stress the importance of understanding the physics of the supernovae, if not only to underpin the current cosmological claims, but also to enable future precision cosmology using SNe Ia. 

Major efforts are also presently undertaken to pursue such research on supernova physics, for example within the 
EU Research and Training 
Network\footnote{www.mpa-garching.mpg.de/$\sim$rtn/.}. Having said this, it is important to make clear that SNe Ia are, in fact, extraordinary accurate as standard candles. While supernova astronomers worry about the details, other cosmologist today are enthusiastically  creative with suggestions on how to observationally determine $w$, using gamma-ray bursts (GRB),
black hole gravitational wave infall, quasar absorption line studies, 
GRB afterglow characteristics, all kinds of gravitational lensing, and more. Some of these suggestions are likely to complement ongoing and future supernova surveys. Most will probably not.

Particular interest was raised concerning gamma-ray bursts, following the discovery of the Ghirlanda-relation \citep{ghirlanda04}. There are several aspects of gamma-ray bursts that immediately make them very interesting {\it if} they prove possible to properly calibrate: They are extremely bright, we know that they exist also at very high redshifts, and the gamma-ray properties are not affected by intervening dust. This has raised a flurry of investigations and recently even a suggestion for a dedicated GRB-cosmology dark energy satellite \citep{lamb05}. However, it may well be that the redshift distribution of GRBs is not as optimal as is the case for SNe Ia. In \citet{mortsell05} we showed that the GRB cosmology is mainly sensitive to the matter density probed at higher redshifts, and not efficient in constraining the properties of the dark energy.

SNe Ia are indeed exceptionally good standard candle candidates. They are bright and show a small dispersion in the Hubble diagram. Seen the other way around, it is SNe Ia that provide the best evidence for a linear Hubble expansion in the local universe \citep[see e.g.,][]{leibundguttammann90,riess96}. Despite the worries voiced above, the theoretical understanding of SNe Ia is considerable, and much better than can be claimed for e.g., gamma-ray bursts. The redshift distribution of SNe Ia is also very favorable for investigations of the dark energy, and the local sample is important to tie the high-z sample to the Hubble diagram.
Moreover, the local supernova sample makes it possible to understand these phenomena in detail, and to directly compare them in different environs.

\subsection{Epilogue}

When the acceleration of the cosmic expansion was first claimed 7 years ago, it was certainly strengthened by the fact that two independent international teams \citep{schmidt98,perlmutter99} reached the same conclusions. The ESSENCE project, as a continuation of the HZT efforts, is today working within the concordance cosmology paradigm.
But even if a detection of new physics, in the form of a $w\not=-1$ measurement, may not be as large a shock for the already perplexed physics community as the initial $\Omega_{\rm X} > 0$ result, it is likely that the competition with the SNLS will prove healthy also this time. And after all, a result where $w< -1$ is still not ruled out.

\section*{Acknowledgments}

I want to thank Jakob J\"onsson for some calculations.
I want to thank the organizers for inviting me to the
{\it 13th General Meeting of the European Physical Society conference, 
Beyond Einstein: Physics for the 21st Century,  
Conference II: Relativity, Matter and Cosmology} and the 
Swedish Research Council for travel grants.
Part of this research was done within the DARK cosmology 
center funded by the Danish National Research Foundation.

% The following bibliography was produced with
%   \bibliographystyle{aa}
%   \bibliography{esapub}
% The results are inserted directly here to simplify
% the demonstration.

\end{document}